\documentclass[aps,prd,twocolumn,superscriptaddress,showpacs,preprintnumbers,amsmath,amssymb]{revtex4-1}

\usepackage{xspace}
\usepackage{url}
\usepackage{xcolor}
\usepackage{graphicx} 
\usepackage{dcolumn}  
\usepackage[compat=1.1.0]{tikz-feynman}
\everymath{\displaystyle}

\newcommand{\gev}{{\rm{\,Ge\kern -0.1em V}}\xspace}
\newcommand{\mev}{{\rm{\,Me\kern -0.1em V}}\xspace}
\newcommand{\gevc}{{{{\rm \,Ge\kern -0.1em V\!/}}c}\xspace}
\newcommand{\mevc}{{{{\rm \,Me\kern -0.1em V\!/}}c}\xspace}
\newcommand{\gevcc}{{{{\rm \,Ge\kern -0.1em V\!/}}c^2}\xspace}
\newcommand{\mevcc}{{{{\rm\,Me\kern -0.1em V\!/}}c^2}\xspace}

\def\KL  {\ensuremath{K^0_{\scriptscriptstyle L}}\xspace}
\def\CP {\ensuremath{C\!P}\xspace}
\def\Bbar{\kern 0.18em\bar{\kern -0.18em B}{}\xspace}
\def\Dbar{\kern 0.18em\bar{\kern -0.18em D}{}\xspace}

\graphicspath{{ps}}

\begin{document}
\vspace*{-3\baselineskip}
\resizebox{!}{3cm}{\includegraphics{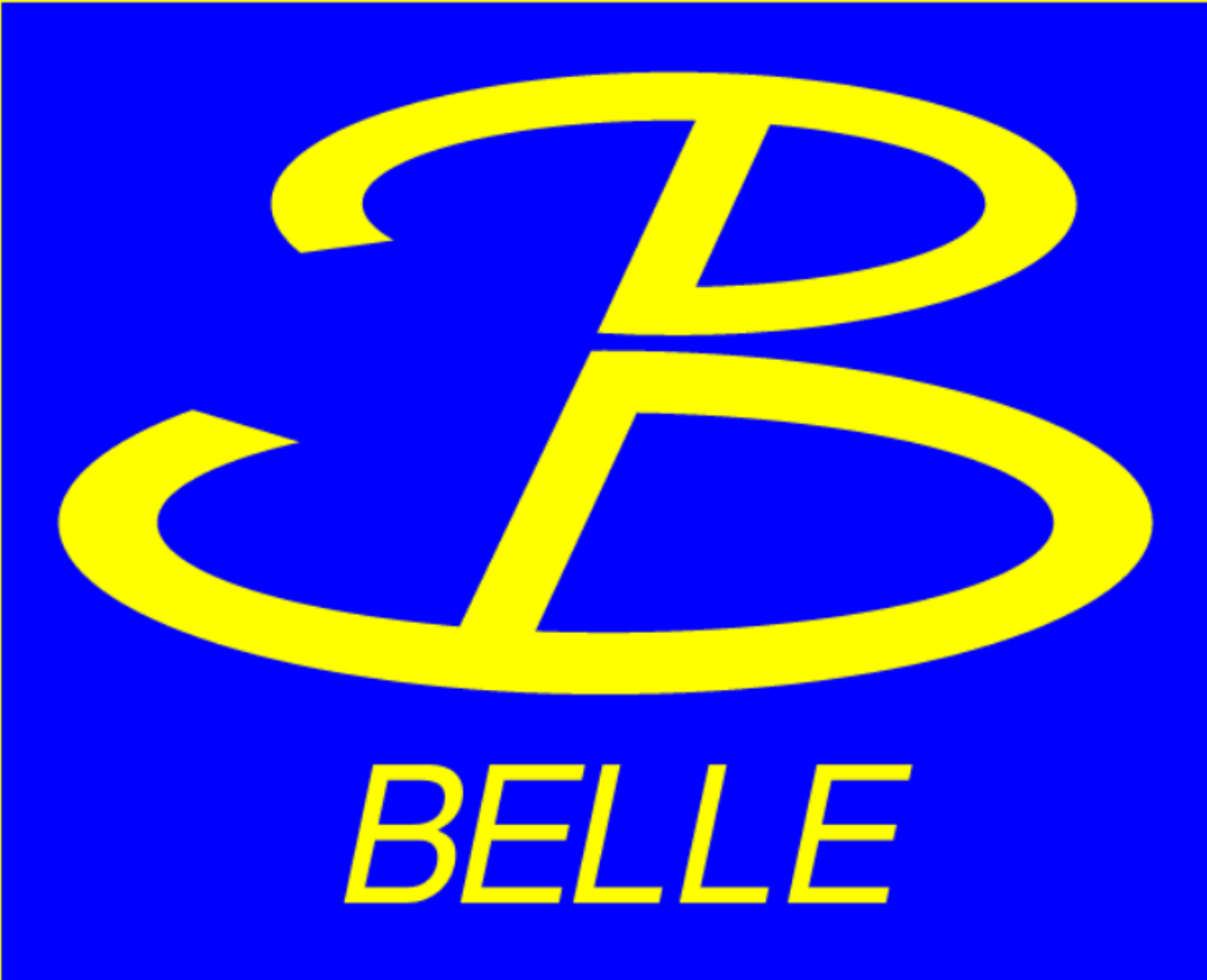}}

\preprint{\vbox{ \hbox{   }
    \hbox{Belle Preprint 2021-17}
    	\hbox{KEK Preprint 2021-15}
}}

\title{ \quad\\[1.0cm] {Measurement of the branching fraction and $\CP$ asymmetry for $B\to\bar{D}^{0} \pi$ decays}}
\noaffiliation
\affiliation{Department of Physics, University of the Basque Country UPV/EHU, 48080 Bilbao}
\affiliation{University of Bonn, 53115 Bonn}
\affiliation{Brookhaven National Laboratory, Upton, New York 11973}
\affiliation{Budker Institute of Nuclear Physics SB RAS, Novosibirsk 630090}
\affiliation{Faculty of Mathematics and Physics, Charles University, 121 16 Prague}
\affiliation{Chonnam National University, Gwangju 61186}
\affiliation{University of Cincinnati, Cincinnati, Ohio 45221}
\affiliation{Deutsches Elektronen--Synchrotron, 22607 Hamburg}
\affiliation{Duke University, Durham, North Carolina 27708}
\affiliation{University of Florida, Gainesville, Florida 32611}
\affiliation{Department of Physics, Fu Jen Catholic University, Taipei 24205}
\affiliation{Key Laboratory of Nuclear Physics and Ion-beam Application (MOE) and Institute of Modern Physics, Fudan University, Shanghai 200443}
\affiliation{Justus-Liebig-Universit\"at Gie\ss{}en, 35392 Gie\ss{}en}
\affiliation{Gifu University, Gifu 501-1193}
\affiliation{SOKENDAI (The Graduate University for Advanced Studies), Hayama 240-0193}
\affiliation{Gyeongsang National University, Jinju 52828}
\affiliation{Department of Physics and Institute of Natural Sciences, Hanyang University, Seoul 04763}
\affiliation{University of Hawaii, Honolulu, Hawaii 96822}
\affiliation{High Energy Accelerator Research Organization (KEK), Tsukuba 305-0801}
\affiliation{J-PARC Branch, KEK Theory Center, High Energy Accelerator Research Organization (KEK), Tsukuba 305-0801}
\affiliation{Higher School of Economics (HSE), Moscow 101000}
\affiliation{Forschungszentrum J\"{u}lich, 52425 J\"{u}lich}
\affiliation{IKERBASQUE, Basque Foundation for Science, 48013 Bilbao}
\affiliation{Indian Institute of Technology Bhubaneswar, Satya Nagar 751007}
\affiliation{Indian Institute of Technology Hyderabad, Telangana 502285}
\affiliation{Indian Institute of Technology Madras, Chennai 600036}
\affiliation{Indiana University, Bloomington, Indiana 47408}
\affiliation{Institute of High Energy Physics, Chinese Academy of Sciences, Beijing 100049}
\affiliation{Institute of High Energy Physics, Vienna 1050}
\affiliation{Institute for High Energy Physics, Protvino 142281}
\affiliation{INFN - Sezione di Napoli, 80126 Napoli}
\affiliation{INFN - Sezione di Torino, 10125 Torino}
\affiliation{Advanced Science Research Center, Japan Atomic Energy Agency, Naka 319-1195}
\affiliation{J. Stefan Institute, 1000 Ljubljana}
\affiliation{Institut f\"ur Experimentelle Teilchenphysik, Karlsruher Institut f\"ur Technologie, 76131 Karlsruhe}
\affiliation{Kavli Institute for the Physics and Mathematics of the Universe (WPI), University of Tokyo, Kashiwa 277-8583}
\affiliation{Department of Physics, Faculty of Science, King Abdulaziz University, Jeddah 21589}
\affiliation{Kitasato University, Sagamihara 252-0373}
\affiliation{Korea Institute of Science and Technology Information, Daejeon 34141}
\affiliation{Korea University, Seoul 02841}
\affiliation{Kyoto Sangyo University, Kyoto 603-8555}
\affiliation{Kyungpook National University, Daegu 41566}
\affiliation{Universit\'{e} Paris-Saclay, CNRS/IN2P3, IJCLab, 91405 Orsay}
\affiliation{P.N. Lebedev Physical Institute of the Russian Academy of Sciences, Moscow 119991}
\affiliation{Liaoning Normal University, Dalian 116029}
\affiliation{Faculty of Mathematics and Physics, University of Ljubljana, 1000 Ljubljana}
\affiliation{Ludwig Maximilians University, 80539 Munich}
\affiliation{Luther College, Decorah, Iowa 52101}
\affiliation{Malaviya National Institute of Technology Jaipur, Jaipur 302017}
\affiliation{Faculty of Chemistry and Chemical Engineering, University of Maribor, 2000 Maribor, Slovenia} 
\affiliation{Max-Planck-Institut f\"ur Physik, 80805 M\"unchen}
\affiliation{School of Physics, University of Melbourne, Victoria 3010}
\affiliation{University of Mississippi, University, Mississippi 38677}
\affiliation{University of Miyazaki, Miyazaki 889-2192}
\affiliation{Moscow Physical Engineering Institute, Moscow 115409}
\affiliation{Graduate School of Science, Nagoya University, Nagoya 464-8602}
\affiliation{Kobayashi-Maskawa Institute, Nagoya University, Nagoya 464-8602}
\affiliation{Universit\`{a} di Napoli Federico II, 80126 Napoli}
\affiliation{Nara Women's University, Nara 630-8506}
\affiliation{National Central University, Chung-li 32054}
\affiliation{National United University, Miao Li 36003}
\affiliation{Department of Physics, National Taiwan University, Taipei 10617}
\affiliation{H. Niewodniczanski Institute of Nuclear Physics, Krakow 31-342}
\affiliation{Nippon Dental University, Niigata 951-8580}
\affiliation{Niigata University, Niigata 950-2181}
\affiliation{Novosibirsk State University, Novosibirsk 630090}
\affiliation{Osaka City University, Osaka 558-8585}
\affiliation{Pacific Northwest National Laboratory, Richland, Washington 99352}
\affiliation{Panjab University, Chandigarh 160014}
\affiliation{Peking University, Beijing 100871}
\affiliation{University of Pittsburgh, Pittsburgh, Pennsylvania 15260}
\affiliation{Punjab Agricultural University, Ludhiana 141004}
\affiliation{Research Center for Nuclear Physics, Osaka University, Osaka 567-0047}
\affiliation{Meson Science Laboratory, Cluster for Pioneering Research, RIKEN, Saitama 351-0198}
\affiliation{Department of Modern Physics and State Key Laboratory of Particle Detection and Electronics, University of Science and Technology of China, Hefei 230026}
\affiliation{Seoul National University, Seoul 08826}
\affiliation{Showa Pharmaceutical University, Tokyo 194-8543}
\affiliation{Soochow University, Suzhou 215006}
\affiliation{Soongsil University, Seoul 06978}
\affiliation{Sungkyunkwan University, Suwon 16419}
\affiliation{School of Physics, University of Sydney, New South Wales 2006}
\affiliation{Department of Physics, Faculty of Science, University of Tabuk, Tabuk 71451}
\affiliation{Tata Institute of Fundamental Research, Mumbai 400005}
\affiliation{Department of Physics, Technische Universit\"at M\"unchen, 85748 Garching}
\affiliation{School of Physics and Astronomy, Tel Aviv University, Tel Aviv 69978}
\affiliation{Department of Physics, Tohoku University, Sendai 980-8578}
\affiliation{Earthquake Research Institute, University of Tokyo, Tokyo 113-0032}
\affiliation{Department of Physics, University of Tokyo, Tokyo 113-0033}
\affiliation{Tokyo Institute of Technology, Tokyo 152-8550}
\affiliation{Tokyo Metropolitan University, Tokyo 192-0397}
\affiliation{Utkal University, Bhubaneswar 751004}
\affiliation{Virginia Polytechnic Institute and State University, Blacksburg, Virginia 24061}
\affiliation{Wayne State University, Detroit, Michigan 48202}
\affiliation{Yamagata University, Yamagata 990-8560}
\affiliation{Yonsei University, Seoul 03722}
  \author{T.~Bloomfield}\affiliation{School of Physics, University of Melbourne, Victoria 3010} 
  \author{M.~E.~Sevior}\affiliation{School of Physics, University of Melbourne, Victoria 3010} 
  \author{I.~Adachi}\affiliation{High Energy Accelerator Research Organization (KEK), Tsukuba 305-0801}\affiliation{SOKENDAI (The Graduate University for Advanced Studies), Hayama 240-0193} 
  \author{H.~Aihara}\affiliation{Department of Physics, University of Tokyo, Tokyo 113-0033} 
  \author{S.~Al~Said}\affiliation{Department of Physics, Faculty of Science, University of Tabuk, Tabuk 71451}\affiliation{Department of Physics, Faculty of Science, King Abdulaziz University, Jeddah 21589} 
  \author{D.~M.~Asner}\affiliation{Brookhaven National Laboratory, Upton, New York 11973} 
  \author{V.~Aulchenko}\affiliation{Budker Institute of Nuclear Physics SB RAS, Novosibirsk 630090}\affiliation{Novosibirsk State University, Novosibirsk 630090} 
  \author{T.~Aushev}\affiliation{Higher School of Economics (HSE), Moscow 101000} 
  \author{R.~Ayad}\affiliation{Department of Physics, Faculty of Science, University of Tabuk, Tabuk 71451} 
  \author{V.~Babu}\affiliation{Deutsches Elektronen--Synchrotron, 22607 Hamburg} 
% ...
\collaboration{The Belle Collaboration}


\begin{abstract} 
We measure the branching fractions and $\CP$ asymmetries for the decays $B^{0}\to \bar{D}^{0}\pi^{0}$ and $B^{+}\to \bar{D}^{0}\pi^{+}$, using a data sample of $772\times 10^{6}$ $B\Bbar$ pairs collected at the $\Upsilon(4S)$ resonance with the Belle detector at the KEKB $e^{+}e^{-}$ collider. The branching fractions obtained and direct $\CP$ asymmetries are 
$\mathcal{B}(B^{0}\to \bar{D}^{0}\pi^{0}) = [2.70 \pm 0.06~ \text{(stat.)} \pm 0.10~ \text{(syst.)}] \times 10^{-4}$, 
$\mathcal{B}(B^{+}\to \bar{D}^{0}\pi^{+}) = [4.53 \pm 0.02~ \text{(stat.)} \pm 0.15~ \text{(syst.)}] \times 10^{-3}$, 
$ {\cal A}_{\CP}(B^{0}\to \bar{D}^{0}\pi^{0})  =  [+0.42 \pm 2.05~ \text{(stat.)} \pm 1.22~ \text{(syst.)}]\%$, and 
$ {\cal A}_{\CP}(B^{+}\to \bar{D}^{0}\pi^{+})  =  [+0.19 \pm 0.36~ \text{(stat.)} \pm 0.57~ \text{(syst.)}]\%$. The measurements of $\mathcal{B}$ are the most precise to date and are in good agreement with previous results, as is the measurement of ${\cal A}_{\CP}(B^{+}\to \bar{D}^{0}\pi^{+})$. The measurement of ${\cal A}_{\CP}$ for $B^{0}\to \bar{D}^{0}\pi^{0}$ is the first for this mode, and the value is consistent with Standard Model expectations.

\end{abstract}

\pacs{13.25.Hw, 12.15.Hh}

\maketitle

\tighten

{\renewcommand{\thefootnote}{\fnsymbol{footnote}}}
\setcounter{footnote}{0}

\section{\label{sec:introduction}Introduction}

The branching fraction (${\cal B}$) of the color-suppressed decay $B^{0}\to\bar{D}^{0}\pi^{0}$~\cite{CC} is measured~\cite{abe,coan,blyth} to be about a factor of four higher than theory predictions made using the ``naive'' factorization model, where final-state interactions (FSIs) are neglected~\cite{Beneke,Naive_factor_Br}. This has led to a number of new theoretical descriptions of the process~\cite{Neubert,Chua,Color_Supp_Hou,Rosner,Deandrea,Chiang,Color_Supp_Chua_Hou,Nonfactor_D0pi0} that include FSI's and also treat isospin-related amplitudes of color-suppressed and color-allowed decays. The $B^{0}\to\bar{D}^{0}\pi^{0}$ process has been shown to have large non-factorizable components~\cite{Nonfactor_D0pi0}, so precise measurements of its properties are valuable in comparing different theoretical models used to describe it. Many of these models predict a substantial strong phase in the final state. A non-vanishing strong phase difference between two amplitudes is necessary to give rise to direct $\CP$ violation~\cite{bigi}. The direct $\CP$-violation parameter, ${\cal A}_{\CP}$, for the $B \to \bar{D}^{0} \pi$  decay is defined as:
\begin{equation} \label{eq:acp}
  {\cal A}_{\CP} = \frac{\Gamma(\Bbar\to D^{0} \pi) - \Gamma(B\to \bar{D}^{0} \pi)}{\Gamma(\Bbar\to D^{0} \pi) + \Gamma(B\to \bar{D}^{0} \pi)},
\end{equation}
where $\Gamma$ is the partial decay width for the corresponding decay.

\begin{figure}
    \begin{tikzpicture}
        \begin{feynman}
            \vertex (a1) {\(\bar b\)};
            \vertex[right=2cm of a1] (a3) [label=90:\small\(V_{cb}\)];
            \vertex[right=2cm of a3] (a5) {\(\bar c\)};
            \vertex[below=1.5cm of a1] (b1) {\(d\)};
            \vertex[below=1.5cm of a5] (b2) {\(d\)};
            \vertex[below=1.0em of a5] (c1) {\(u\)};
            \vertex[above=1.0em of b2] (c3) {\(\bar d\)};
            \vertex at ($(c1)!0.5!(c3) - (1cm, 0)$) (c2) [label={[label distance=0.2em]270:\small\(V^{*}_{ud}\)}];
            \diagram* {
            {[edges=fermion]
            (a5) -- (a3) -- (a1),
            },
            (b1) -- [fermion] (b2),
            (c3) -- [fermion, out=180, in=-60] (c2) -- [fermion, out=60, in=180] (c1),
            (a3) -- [boson, edge label=\(W^{+}\), bend right] (c2),
            };
            \draw [decoration={brace}, decorate] (b1.south west) -- (a1.north west)
            node [pos=0.5, left] {\(B^{0}\)};
            \draw [decoration={brace}, decorate] (a5.north east) -- (c1.south east)
            node [pos=0.5, right] {\(\bar{D}^{0}\)};
            \draw [decoration={brace}, decorate] (c3.north east) -- (b2.south east)
            node [pos=0.5, right] {\(\pi^{0}\)};
            \node[left= 1cm of a1] {a)};
        \end{feynman}
    \end{tikzpicture}
    \begin{tikzpicture}
        \begin{feynman}
            \vertex (a1) {\(\overline b\)};
            \vertex[right=2cm of a1] (a3) [label=90:\small\(V_{cb}\)];
            \vertex[below=2.5em of a3] (b3) [label=270:\small\(V^{*}_{ud}\)];
            \vertex[right=2cm of a3] (a5) {\(\overline c\)};
            \vertex[below left =2.2cm and 0.2cm of a5] (c1) {\(\overline d/\overline u\)};
            \vertex[above=2.0em of c1] (c3) {\(d/u\)};
            \vertex[below=2.5em of a1] (b1) {\(d\)};
            \vertex[below=2.5em of a5] (b2) {\(u\)};
            \vertex at ($(c1)!0.5!(c3) - (1.5cm, 0)$) (c2);
            \diagram* {
            {[edges=anti fermion]
            (a1) -- (a3) -- (a5),
            },
            (c1) -- [fermion, out=180, in=-60] (c2) -- [fermion, out=60, in=180] (c3),
            (a3) -- [boson, edge label=\small\(W\), bend left] (b3),
            (b1) -- [fermion] (b3),
            (b3) -- [fermion] (b2),
            };
            \node[left= 1cm of a1] {b)};
            \draw [decoration={brace}, decorate] (b1.south west) -- (a1.north west)
            node [pos=0.5, left] {\(B^{0}\)};
            \draw [decoration={brace}, decorate] (a5.north east) -- (b2.south east)
            node [pos=0.5, right] {\(\bar{D}^{0}\)};
            \draw [decoration={brace}, decorate] (c3.north east) -- (c1.south east)
            node [pos=0.5, right] {\(\pi^{0}\)};
        \end{feynman}
    \end{tikzpicture}
    \begin{tikzpicture}
        \begin{feynman}
            \vertex (a1) {\(b\)};
            \vertex[right=2cm of a1] (a3) [label=90:\small\(V^{*}_{ub}\)];
            \vertex[right=2cm of a3] (a5) {\(u\)};
            \vertex[below=1.5cm of a1] (b1) {\(\bar d\)};
            \vertex[below=1.5cm of a5] (b2) {\(\bar d\)};
            \vertex[below=1.0em of a5] (c1) {\(\bar c\)};
            \vertex[above=1.0em of b2] (c3) {\(d\)};
            \vertex at ($(c1)!0.5!(c3) - (1cm, 0)$) (c2) [label={[label distance=0.2em]270:\small\(V_{cd}\)}];
            \diagram* {
            {[edges=fermion]
            (a1) -- (a3) -- (a5),
            },
            (b2) -- [fermion] (b1),
            (c3) -- [anti fermion, out=180, in=-60] (c2) -- [anti fermion, out=60, in=180] (c1),
            (a3) -- [boson, edge label=\(W^{-}\), bend right] (c2),
            };
            \draw [decoration={brace}, decorate] (b1.south west) -- (a1.north west)
            node [pos=0.5, left] {\(\bar{B}^{0}\)};
            \draw [decoration={brace}, decorate] (a5.north east) -- (c1.south east)
            node [pos=0.5, right] {\(\bar{D}^{0}\)};
            \draw [decoration={brace}, decorate] (c3.north east) -- (b2.south east)
            node [pos=0.5, right] {\(\pi^{0}\)};
             \node[left= 1cm of a1] {c)};
        \end{feynman}
    \end{tikzpicture}
    \caption{Tree-level Feynman diagrams for a) color-suppressed $B^{0}\to \bar{D}^{0}\pi^{0}$ decay, b) W exchange diagram, and c) color-suppressed and doubly Cabibbo-suppressed $\bar{B}^{0}\to \bar{D}^{0}\pi^{0}$ decay.}
    \label{figfeynman}
\end{figure}
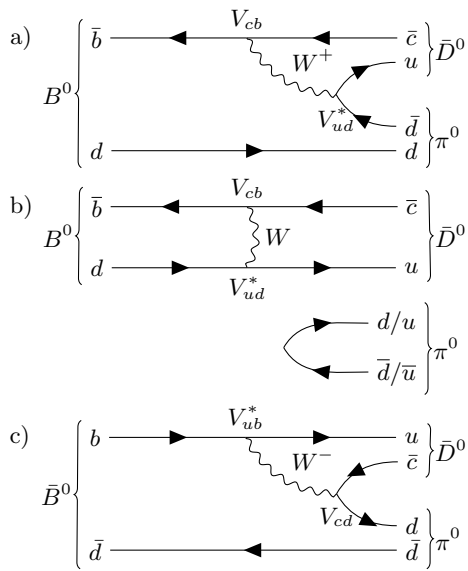

In the Standard Model (SM), $B^{0}\to\bar{D}^{0}\pi^{0}$ transitions proceed mainly via the tree-level diagram of Fig.\ref{figfeynman}a. An exchange diagram (Fig.\ref{figfeynman}b) with the same CKM factors is also present, but, due to OZI suppression, is expected to have a much smaller amplitude. As such, direct $\CP$ violation in this mode is expected to be small, even in the presence of a strong phase difference from FSI. Measurements of notable $\CP$ violation in this decay would be of significant interest and could hint at contributions from Beyond-the-Standard-Model (BSM) physics diagrams.   
Recently the BaBar and Belle collaborations performed time-dependent $\CP$-violation analyses of the related modes $\bar{B}^{0}\to D^{(*)}_{CP}h^{0}$, where $h \in {\pi^{0},\eta,\omega}$ and $D^{(*)}_{CP}$ refers to $D$ or $D^{*}$ in a $CP$ eigenstate~\cite{bellebabar}. They measure the $\CP$-violation parameters $C (=-{\cal A}_{CP})$, ${\cal S}_{CP}$ and $\phi_{1}$\cite{alpha}, and obtain $C(B^{0}\to\bar{D}^{*0}h^{0}) = (-2 \pm 8)\times10^{-2}$. This value is consistent with the expectation of small ${\cal A}_{CP}$ for $B^{0}\to\bar{D}^{0}\pi^{0}$. However, this result does not exclude larger values up to 0.1, which would be much larger than SM predictions. 

A high precision measurement of ${\cal B}$ and ${\cal A}_{CP}$ for  $B^{0}\to\bar{D}^{0}\pi^{0}$ has further utility in addition to comparison to theoretical predictions, as it is a common control mode for use in rare charmless $B$ decays with a $\pi^{0}$. The precise measurement of properties of a control mode is important to provide validation and refinement of analysis techniques.

In this paper, we present new measurements of $B^{0}\to\bar{D}^{0}\pi^{0}$ using the full data sample of $(772\pm10.6)\times 10^{6} \bar{B}B$ pairs ($711~\text{fb}^{-1}$) collected with the Belle detector at the KEKB asymmetric-energy $e^+e^-$ ($3.5\gev$ on $8.0\gev$) collider~\cite{KEKB} operating near the $\Upsilon(4S)$ resonance. We also present corresponding measurements of $B^{+}\to\bar{D}^{0}\pi^{+}$ decays, which proceed via a simple spectator diagram with no color suppression.


\section{\label{sec:detector}Belle Detector}
The Belle detector~\cite{Belle} is a large-solid-angle magnetic spectrometer that consists of a silicon vertex detector (SVD), a 50-layer central drift chamber (CDC), an array of aerogel threshold Cherenkov counters (ACC), a barrel-like arrangement of time-of-flight scintillation counters (TOF), and an electromagnetic calorimeter (ECL) consisting of CsI(Tl) crystals.
All these detector components are located inside a superconducting solenoid coil that provides a 1.5\,T magnetic field.
An iron flux-return located outside of the coil is instrumented with resistive plate chambers to detect $\KL$ mesons and to identify muons.
Two inner detector configurations were used: a 2.0 cm beam-pipe
and a 3-layer SVD (SVD1) were used for the first sample of $152 \times 10^6$ $B\Bbar$ pairs, while a 1.5 cm beam-pipe, a 4-layer SVD (SVD2), and small cells in the inner layers of the CDC were used to record the remaining $620 \times 10^6$ $B\Bbar$ pairs~\cite{svd2}.

We reconstruct  $B^{0}\to \bar{D}^{0}\pi^{0}$ candidates from the subsequent decays of the $\bar{D}^{0}$ and the $\pi^{0}$ mesons. We employ two reconstruction modes:  $B^{0}\to\bar{D}^{0}(\to K^{+}\pi^{-})\pi^{0}$ ($B_{2b}$) and $B^{0}\to\bar{D}^{0}(\to K^{+}\pi^{-} \pi^{0})\pi^{0}$ ($B_{3b}$).  The $\pi^{0}$ mesons are reconstructed from their decay to two photons.

The flavor of the neutral $B$-meson ($B^{0}$ or $\bar{B}^{0}$) is determined by the charge of the reconstructed kaon ($K^{+}$ or $K^{-}$). This method of flavor tagging is not perfect, as the wrong-sign doubly Cabibbo-suppressed decays (DCS) $\bar{B}^{0}\to \bar{D}^{0} \pi^{0}$ (Fig.\ref{figfeynman}c) will result in wrongly tagged flavor. The same effect occurs with charm DCS decays $D^{0}\to K^{+} \pi^{-}$ and $D^{0}\to K^{+} \pi^{-} \pi^{0}$, and charm mixing, although the charm mixing effect is negligibly smaller. These effects can be calculated using the ratio of wrong-sign (Cabibbo-suppressed) to right-sign (Cabibbo-favored) decay rates:
\begin{equation} \label{eq:WS_RS}
    R=\frac{\cal{B}_{WS}}{\cal{B}_{RS}},
\end{equation}
where
$R(\bar{B}^{0}\to \bar{D}^{0} \pi^{0})$ has not been measured and thus is approximated with $R({B}^{0}\to {D}^{+} \pi^{-})=(2.92\pm0.38\pm0.31)\times 10^{-4}$\cite{Das:2010be}, while $R(D^{0}\to K^{+} \pi^{-})=(3.79\pm0.18)\times 10^{-3}$ and $R(D^{0}\to K^{+} \pi^{-} \pi^{0})=(2.12\pm0.07)\times 10^{-3}$ are taken from the PDG~\cite{PDG_chi_d}. These effects lead to the true value of $\cal{B}$ being $(0.314\pm 0.008)\%$ lower than the measured value, and the true ${\cal A}_{CP}$ being $(3.05\pm 0.08)\times 10^{-5}$ higher than the measured value. In the case of $\cal{B}$ this is corrected for; however, for ${\cal A}_{CP}$ the correction is significantly smaller than our uncertainty and it is neglected.

Photon candidates are mainly taken from clusters in the ECL but additionally are reconstructed from $e^+ e^-$ pairs resulting from photon conversion in the inner detector. Photons from $\pi^{0}$ decay must have an energy greater than $50$ $(100) \mev$ in the barrel (endcap) region of the ECL.
The invariant mass of the two-photon combination must lie in the range $104\mevcc < M_{\gamma \gamma} < 165\mevcc$, corresponding to $\pm 3 \sigma$ around the nominal $\pi^{0}$ mass~\cite{PDG_chi_d}. We subsequently perform a mass-constrained fit with the requirement $\chi^{2} < 50$.

Charged tracks originating from a $B$ decay are required to have a distance-of-closest-approach with respect to the interaction point of less than 4.0 cm along the $z$-axis (the direction opposite the positron beam), and of less than 0.3 cm in the plane transverse to the $z$-axis.
Charged kaons and pions are identified using information from the CDC, ACC, and TOF detectors. This information is combined to form a $K-\pi$ likelihood ratio $R_{K/\pi} = L_K/(L_K + L_\pi)$, where $L_{K}(L_{\pi})$ is the likelihood of the track being a kaon (pion). Track candidates with $R_{K/\pi} > 0.6$ $(< 0.4)$ are classified as kaons (pions). The typical kaon (pion) identification efficiency is
83\% (88\%), with a pion (kaon) misidentification probability of 7\% (11\%).

Two kinematic variables are used to distinguish signal from background: the beam-energy-constrained mass, $M_{\rm bc} \equiv \sqrt{E_{\text{beam}}^{2} - |\vec{p}_{B}|^{2}c^2}/c^2$, and the energy difference $\Delta E \equiv E_{B} - E_\text{beam}$.
Here, $\vec{p}_{B}$ and $E_{B}$ are the momentum and energy, respectively, of the $B$-meson candidate evaluated in the center-of-mass (CM) frame, and $E_{\text{beam}}$ is the beam energy in the CM frame.

Due to energy leakage in the ECL, the reconstructed $\pi^{0}$ energy is typically lower than its true value. To compensate for this, we rescale the reconstructed $\pi^{0}$ momentum to give $ E_{\pi^{0}}\equiv E_{\rm beam} - E_{D}$, specifically:

\begin{align}
\label{eq:MbcCorr}
\vec{p}_{\pi,\text{corr}}  &=\vec{p}_{\pi} \times \frac{\sqrt{(E_{\text{beam}}-E_{D^{0}} )^{2}-M_{\pi^0}^{2}c^{4}}}{|p_{\pi}|c}.
\end{align}
Using this we calculate a new $B$-meson momentum, $\vec{p}_{B,\text{corr}}$, then calculate a corrected $M_{\rm bc}$ (from now on simply referred to as $M_{\rm bc}$),

\begin{equation} \label{eq:MBC}
M_{\rm bc} \equiv \frac{\sqrt{E_{\text{beam}}^{2} - |\vec{p}_{B,\text{corr}}|^{2}c^2}}{c^{2}}.
\end{equation}
Rescaling $M_{\rm bc}$ in this way improves the mass resolution and removes some correlations between $M_{\rm bc}$ and $\Delta E$. This procedure is only applied to the $\pi^{0}$ that is the direct daughter of the $B^{0}$.

All candidates satisfying $M_{\text{bc}} > 5.25 \gevcc$ and $-0.2\gev < \Delta E < 0.2\gev$ are retained for further analysis.
We find that $16\%$ $(47\%)$ of events have more than one $B^{0}$ candidate in the $B_{2b}$ ($B_{3b}$) reconstruction modes.
In these cases, we select one of the reconstructed $B^{0}$ mesons based on the mass difference $\Delta m(X) = m_{\text{PDG}} (X)-M(X)$, where $m_{\text{PDG}} (X)$ is the mass reported by the Particle Data Group (PDG)~\cite{PDG_chi_d} for particle $X$, and $M(X)$ is the reconstructed mass. The best candidate is selected as the $B^0$ or $B^+$ that minimizes $\Delta m(D^{0})$. If there are multiple candidates with the same minimal $\Delta m(D^{0})$, the one that minimizes $\Delta m(\pi^{0})$ is selected. Monte Carlo simulation (MC) studies show that this procedure selects the correct $B^{0}$ in 96\% (86\%) of cases for the $B_{2b}$ ($B_{3b}$) samples.   

\section{\label{sec:backgrounds}Belle detector and signal selection}

Backgrounds to our signal are studied using MC simulation. These simulations use EvtGen~\cite{evtgen} and PYTHIA~\cite{pythia} to generate the physics interactions at the quark level, and employ GEANT3~\cite{GEANT} to simulate the detector response.

The largest background arises from $e^{+} e^{-} \to q \bar{q} \; (q \in \{u, d, s, c\})$ continuum events. A neural network~\cite{neurobayes} is
used to distinguish the spherical $B\bar{B}$ signal from the
jet-like continuum background. It combines the following five observables based on the event topology: a Fisher
discriminant formed from 17 modified Fox-Wolfram moments~\cite{SFW}; the cosine of the angle between the $B$-meson
candidate direction and the beam axis; the cosine of the
angle between the thrust axis~\cite{thrust} of the $B$-meson candidate and that of the rest of the event (all of these quantities
being calculated in the CM frame); the separation along
the $z$-axis between the vertex of the $B$-meson candidate
and that of the remaining tracks in the event; and the tagging quality variable from a $B$-meson flavor-tagging algorithm~\cite{kukuno}.
The training and optimization of the neural network are performed with signal and continuum MC samples. These are divided into five training samples and one verification sample.
The output of the neural net ($C_{\rm NN}$) has a range of $(-1,1)$, with $1$ being the most signal-like and $-1$ being the most background-like.

In order to maximally use $C_{\rm NN}$ information, we impose only a loose requirement on $C_{\rm NN}$ and use $C_{\rm NN}$ as a variable in the fit. We require $C_{\rm NN}>-0.05$ for both the $B_{2b}$ and $B_{3b}$ modes. This results in 86\% background reduction and 87\% signal efficiency.
To facilitate modelling $C_{\rm NN}$ analytically with Gaussian functions, we transform it to an alternative variable $C_{\rm NN}'$ via the formula
\begin{equation}\label{CNN}
   C_{\rm NN}'=\text{log} \left(\frac{C_{\rm NN}-C_{\rm NN}^{\text{min}}}{C_{\rm NN}^{\text{max}}-C_{\rm NN} } \right),
  \end{equation}
where $C_{\rm NN}^{\text{min}}$ is the minimum value of $-0.05$, and $C_{\rm NN}^{\text{max}}$ is the maximum value of $C_{\rm NN}$ obtained from the signal MC sample used to verify the training.

There is a significant background arising from $b\to c$ transitions, which we refer to as ``generic $B$'' decays. The main components of the generic $B$ background are incorrectly assigned tracks, combinatorial backgrounds, $B^{0} \to \bar{D}^{0} \rho^{0}$, and $B^{0} \to \bar{D}^{0*}\pi^{0}$ with either $\bar{D}^{0*} \to \bar{D}^{0}\gamma$ or $\bar{D}^{0*} \to \bar{D}^{0}\pi^{0}$. These are investigated with MC simulations of $B\bar{B}$ decays. To reduce this background, signal candidates are selected within $\pm3$ standard deviations of the mean values for $M_{\rm bc}$, $\Delta E$, and the reconstructed $\pi^{0}$ and $D^{0}$ mass distributions. Low final-state momentum events are excluded with selection criteria on the lab-frame momentum of the final-state particles: $P(K^{\pm})$, $P(\pi^{\pm})$, $P(\pi^{0}_{B^{0}})$, and $P(\pi^{0}_{\bar{D}^{0}})$. These requirements are listed in Table \ref{selections}.

\begin{table}[htb]
\caption{ Requirements on kinematic variables employed in the reconstruction of $B^{0}\to \bar{D}^{0} \pi^{0}$ decays, to minimize generic $B$-decay background. The subscript identifies the origin of the particle, and the $\bar{D}^{0}$ decay mode is shown in square brackets.}
\label{selections}
\begin{tabular}
{@{\hspace{0.5cm}}l@{\hspace{0.5cm}}||@{\hspace{0.5cm}}c@{\hspace{0.5cm}}}
\hline \hline
 Variable  & Selected Range  \\
\hline
$M_{\rm bc}$ & $5.253 \text{ -- } 5.288 \gevcc$ \\
$\Delta E$ & $-0.2 \text{ -- } 0.2 \gev$ \\
$M(D^{0})$  & $ 1.841 \text{ -- } 1.882 \gevcc$ \\
$M(\pi^{0}_{B^{0}})[K^{+}\pi^{-}]$  & $104.1 \text{ -- } 163.1 \mevcc$ \\
$M(\pi^{0}_{B^{0}})[K^{+}\pi^{-}\pi^{0}]$  & $ 105.8 \text{ -- } 164.4\mevcc$ \\
$M(\pi^{0}_{\bar{D}^{0}})[K^{+}\pi^{-}\pi^{0}]$  & $107.9 \text{ -- } 162.3 \mevcc$ \\
$P(K^{\pm})$ & $ 0.3 \text{ -- } 3.5 \gevc$\\
$P(\pi^{\pm})$ & $ 0.3 \text{ -- } 3.5 \gevc$ \\
$P(\pi^{0}_{B^{0}})$ & $ 1.5 \text{ -- } 3.5 \gevc$\\
$P(\pi^{0}_{\bar{D}^{0}})$ & $ 0.2 \text{ -- } 3.5 \gevc$\\
\hline \hline
\end{tabular}
\end{table}

There is also a very small background component from $b\to u$ and $b\to s$ transitions that consists mainly of combinatorial background, with some non-resonant $B$ decays to the same final states ($K^{+}\pi^{-}\pi^{0}$, $K^{+}\pi^{-}\pi^{0} \pi^{0}$, $K^{+}\pi^{-}\pi^{+}$ and $K^{+}\pi^{-}\pi^{+} \pi^{0}$). These are studied using a large MC sample corresponding to 50 times the number of $B\bar{B}$ events recorded by Belle.
We refer to these background events as ``rare.'' The yield of these rare events is fixed when fitting for the signal yield based on the most recent branching fractions from the PDG~\cite{PDG_chi_d}.

After all selections have been made for $B^{0}\to \bar{D}^{0} \pi^{0}$, the reconstruction efficiencies are ($27.53\pm0.04$)\% for the $B_{2b}$ mode and ($9.43\pm 0.02$)\% for the $B_{3b}$ mode. Including intermediate branching fractions, the overall efficiencies are ($1.09\pm0.01$)\% and ($1.36\pm0.01$)\%, respectively.

For $B^{+}\to \bar{D}^{0} \pi^{+}$, the reconstruction efficiency after all selections is calculated to be ($33.08\pm0.04$)\% for the $B_{2b}$ mode (($1.31\pm0.05)$\% including intermediate branching fractions) and ($9.05\pm 0.02$)\% for the $B_{3b}$ mode (($1.30\pm0.05)$\% including intermediate branching fractions).

\section{\label{sec:fitting}Fitting strategy}

The signal yield and ${\cal A}_{\CP}$ are extracted via an unbinned extended maximum-likelihood fit to the variables $M_{\rm bc}$, $\Delta{} E$, and $C_{\rm NN}'$. There are four categories of events fitted: $B^{0}\to \bar{D}^{0} \pi^{0}$ or $B^{+}\to \bar{D}^{0} \pi^{+}$ signal events ($s$), continuum events ($c$), generic $B\bar{B}$ events ($b$), and rare $B$-decay backgrounds ($r$). These events are described by probability density functions (PDFs) denoted as $P^{s}$, $P^{c}$, $P^{b}$, and $P^{r}$, respectively.
Separate PDFs are constructed for the $B_{2b}$ and $B_{3b}$ reconstruction modes, which are fitted as two separate data sets. The data are further divided into events tagged as $B^{0}$ and $\bar{B}^{0}$, defined as having flavor $q=+1$ and $q=-1$, respectively, based on the charge of the kaon. 

The physics parameters are determined via a simultaneous fit to the four data sets. The total likelihood is given by
\begin{align}
\label{eq:likelihood}
  {\cal L} & =  \frac{e^{-\sum_{j}N^{j}}}{ \prod_{q,d} N_{q,d}!}
  \times \prod_{q,d} \nonumber \\
  & \left[ \prod_{i=1}^{N_{q,d}} 
   \left( \sum_{j}f^{j}_{d}N^{j}P^{j}_{q,d}\left(M_{\text{bc}}^{i},\Delta E^{i},C_{\rm NN}'^{i},q \right) \right) \right].
\end{align}
where  $N_{q,d}$ is the number of events with flavor tag $q$ for the data set $d$ ($d \in {B_{2b}, B_{3b}}$), and $N^{j}$ is the number of events in the $j^{\rm th}$ category ($j \in {s,c,b,r}$) contributing to the total yield.
The fraction of events in the data set $d$ for category $j$ is $f^j_{d}$, with $f^{j}_{3b} = 1 - f^{j}_{2b}$.
The PDF $P^j_{q,d}$ corresponds to the $j^{\rm th}$ category in the $d$ data set for flavor $q$, measured at $M_{\rm bc}^{i}$, $\Delta E^{i}$, and $C_{\rm NN}'^{i}$ for the $i^{\rm th}$ event.

The PDF for each component is given by:
\begin{align} \label{eq:pdf} 
  P^{j}_{q,d}(M_{\text{bc}},\Delta E,C_{\rm NN}',q) =& \nonumber \\
  \left(\frac{1 - q\times {\cal A}^{j}_{\CP}}{2}\right) &\times P^{j}_{d}(M_{\text{bc}},\Delta E,C_{\rm NN}'),
\end{align}
The model accounts for a possible direct $\CP$ asymmetry, ${\cal A}^{j}_{\CP}$, and the fractions of signal and backgrounds expected in each reconstruction mode $B_{2b}$ ($B_{3b}$). 
In Eq.~\eqref{eq:likelihood}, the fraction $f^{s}_{d}$~ is determined via MC studies of the $B_{2b}$ and $B_{3b}$ modes and is fixed in the fit to data, and ${\cal A}^{j}_{\CP} (j \in {c,b,r})$ is fixed based on studies of detector bias using sideband data (see Section \ref{sec:control}). 

The 20 free parameters in the fit are: the number of signal events $N^{s}$, signal asymmetry $A^{s}_{\CP}$, the number of continuum events ($N^{c}$) and generic $B$-decay events ($N^{b}$), fractions of backgrounds expected in each reconstruction mode $f^j_{d}$ ($j \in {c,b,r}$), shape parameters of the continuum $M_{\rm bc}$ and $\Delta E$ PDFs, and the mean and width of the $B_{2b}$ signal $M_{\rm bc}$ and $\Delta E$ PDFs. The number of rare background events ($N^{r}$) is fixed to that expected from MC studies. The effects of these assumptions are included in the systematic uncertainties.

The PDFs used for the $M_{\rm bc}$ and $\Delta E$ distributions for the various event types are as follows.
\begin{itemize}
  \item	Signal: for the $B_{2b}$ mode, the $M_{\rm bc}$ PDF is a Crystal Ball function~\cite{CrystalBall}, while the $\Delta E$ PDF is the sum of a Crystal Ball function and a Gaussian with the same mean. The Gaussian component is small and included to handle the tails of the distribution. For the $B_{3b}$ mode, there is a strong correlation between $M_{\rm bc}$ and $\Delta E$, and no analytic 2D PDF could be found to fit the data satisfactorily. Instead a 2D kernel density estimation (KEST) PDF~\cite{Kernel} is used.
  
  \item Generic $B$-decay background: similarly to $B_{3b}$  signal, there exist complex correlations between $M_{\rm bc}$ and $\Delta E$, so a 2D KEST PDF obtained from MC simulations in both modes.

  \item Continuum background: $M_{\rm bc}$ is fitted as an ARGUS function~\cite{argus}, and $\Delta E$ as a 3rd-order Chebyshev polynomial, in both modes.
    
  \item Rare $B$-decay background: as with generic $B$ and $B_{3b}$ signal, the $M_{\rm bc}$ and $\Delta E$ distributions for rare $B$-decay background are modelled with a 2D KEST PDF in both modes. This PDF is determined using MC simulations corresponding to 50 times the luminosity of the Belle data set.
\end{itemize}

To fit $C_{\rm NN}'$, three summed Gaussians are used for all components except continuum background, which employed two summed Gaussians. The PDFs for all event types are summarized in Table \ref{SignalPDFS_table}.

\begin{table}[htb]
\caption{ Functional forms for PDFs employed by the different event categories for fits.}
\label{SignalPDFS_table}
\begin{tabular}
{@{\hspace{0.2cm}}c@{\hspace{0.2cm}}||@{\hspace{0.2cm}}c@{\hspace{0.2cm}}@{\hspace{0.2cm}}c@{\hspace{0.2cm}}@{\hspace{0.2cm}}c@{\hspace{0.2cm}}}
\hline \hline
Category  & $M_{\rm bc}$ & $\Delta E$ & $C_{\rm NN}'$  \\
\hline
Signal $B_{2b}$ & Crystal Ball & Crystal Ball &	3 Gaussians  \\
                &         &  + Gaussian &	  \\ \hline
Signal $B_{3b}$ & \multicolumn{2}{c}{2D KEST PDF}   & 3 Gaussians  \\ \hline
Generic $B$ & \multicolumn{2}{c}{2D KEST PDF}     & 3 Gaussians  \\ \hline
Continuum &	ARGUS & $3^{\rm rd}$ Order  & 2 Gaussians  \\
                &         &  Cheby. Poly. &	  \\ \hline
Rare $B$	& \multicolumn{2}{c}{2D KEST PDF}	 & 3 Gaussians \\
\hline \hline
\end{tabular}
\end{table}

The fitting procedure and accuracy of the various PDF models are extensively investigated using MC `pseudoexperiments'.
In these studies, the signal and rare $B$ background events are selected from large samples of simulated events.
Events for $e^{+} e^{-} \to q \bar{q}$ and generic $B$-decay are generated from their respective PDF shapes. For $B^{0} \to \bar{D}^{0} \pi^{0}$, we observe a small bias of $(+0.6 \pm 0.3)$\% in signal yield, and $(+0.04 \pm 0.05)$\% in ${\cal A}_{CP}$. We correct for this bias in our final measurements and include a corresponding systematic uncertainty for it. No significant bias is observed for $B^{+} \to \bar{D}^{0} \pi^{+}$, i.e., only $(+0.06 \pm 0.16)$\% in signal yield and $(-0.02 \pm 0.02)$\% in ${\cal A}_{CP}$.

\section{\label{sec:control}$B^{+} \to \bar{D}^{0} \pi^{+}$}

We first select and fit a sample of $B^{+}\to \bar{D}^{0} \pi^{+}$ decays. This sample is not color-suppressed and thus has much larger statistics and lower background than the sample of $B^{0}\to\bar{D}^{0} \pi^{0}$ decays. As well as ensuring the fitted ${\cal B}$ and ${\cal A}_{CP}$ are consistent with existing measurements, this mode is used to obtain calibration factors for the fixed shape parameters of the PDFs used to fit $B^{0} \to \bar{D}^{0} \pi^{0}$ decays, to account for any differences between MC and data. In addition, this mode provides a data-driven estimation of the systematic uncertainty associated with the ${\cal A}_{CP}$ correction for a detection asymmetry (discussed below).

To account for potential differences in the distribution of fitting variables between MC and data, additional parameters (calibration factors) are included in the fit to enable small adjustments in the fitted PDF shapes. These calibration factors are applied as mean shifts and width factors to the $C_{\text{NN}}'$ Gaussians.

To account for small differences between MC and data in the width of the $\Delta E$ distribution, the 2D KEST PDF for $M_{\rm bc}$ and $\Delta E$ in $B_{3b}$ is modified slightly. This is done by modifying each $\Delta E$ data point in the MC data set with a random shift based on a Gaussian distribution with a mean of 0 GeV and a width of 7 MeV (chosen after testing with a range of widths) and generating a new KEST PDF from this modified data sample.

\begin{figure*}[htb]
  \includegraphics[width=1.0\textwidth]{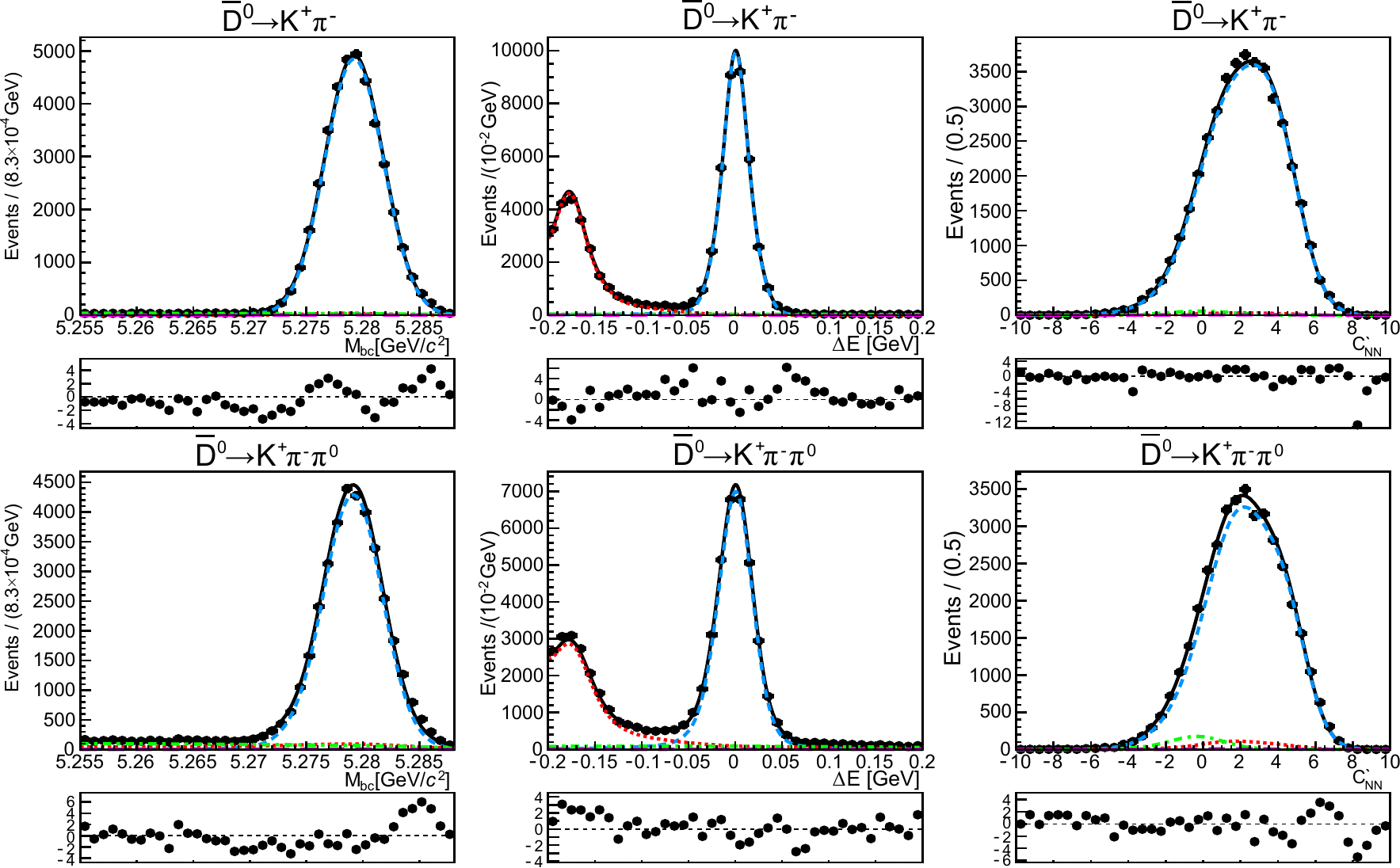}
\caption{Projections of the $B^{+}\to \bar{D}^{0} \pi^{+}$ fit results into the signal region ($5.275<M_{\rm bc} <5.285 \gev$, $-0.05< \Delta E<0.05 \gev$, $-1< C_{\rm NN}' <6$ ) for $M_{\rm bc}$ (left), $\Delta E$ (middle) and $C_{\rm NN}'$ (right) split into the $B_{2b}$ mode (top) and $B_{3b}$ (bottom). The blue short-dashed curve shows the signal PDF, red dotted curve shows the $B\bar{B}$ background PDF, green dash-dotted curve shows the continuum background PDF, pink long-dashed curve shows the (almost negligible) rare background PDF, black line is the fit result, points are data. Also shown underneath each graph is the residual pulls between the data points and fitted PDF.}
\label{yield_Dpi}
\end{figure*}

Fits to the $B^{+}\to\bar{D}^{0} \pi^{+}$ sample are performed to determine the signal yield, ${\cal A}_{CP,\text{raw}}$, the continuum background yield, the generic $B$-decay background yield, and the calibration factors. The rare $B$-decay background yield is fixed to the value expected from MC simulations. From the fit we obtain  $N_{\rm sig} = 84537 \pm 306$ and ${\cal A}_{CP,\text{raw}}=(1.97 \pm 0.36)\%$. The uncertainties listed are statistical. ${\cal A}_{CP,\text{raw}}$ is the output of the fit without a correction to account for sources of bias. Figure~\ref{yield_Dpi} shows the fits to data in $M_{\text{bc}}$, $\Delta E$ and $C_{\rm NN}'$.

To account for possible bias in ${\cal A}_{CP}$, we perform an analysis over a ``sideband'' region of data, defined as $0.1\gev < \Delta E < 0.4 \gev$ and $5.255 \gevcc < M_{\rm bc} < 5.27 \gevcc$. This region consists almost entirely of continuum events and has an expected ${\cal A}_{CP}$ of zero. Counting the number of events in this region we find ${\cal A}_{CP,\text{sideband}} =(1.78 \pm 0.38)\%$. We subtract this value from ${\cal A}_{CP,\text{raw}}$ to correct for the detection asymmetry bias.

The branching fraction is calculated as:
\begin{equation} \label{eq:Br_recipe}
\mathcal{B} = \frac{N_{\rm sig}}{N_{B^{\pm{}}}} \times \text{mean} \left( \frac{f^s_{2b}}{\epsilon_{2b}},\frac{f^s_{3b}}{\epsilon_{3b}} \right),  \\
\end{equation}
where $N_{B^{\pm{}}}$ is the number of charged $B$-mesons in the data set, based on the PDG average value of ${\cal B} (\Upsilon(4S) \to {B}^{+}B^{-})=(51.4\pm0.6\%)$\cite{PDG_chi_d}; $f^s_d$ is the fraction of signal events in data set $d = 2b\text{ or }3b$ ($f^s_{2b} = 0.51$, $f^s_{3b} = 0.49$); and $\epsilon_d$ is the product of the reconstruction efficiency, the $\bar{D}^{0}$ branching fraction $\mathcal{B}_{d}$, and small corrections for particle identification (PID), and charged track and $\pi^0$ reconstruction efficiencies (see Section \ref{systematics}), for mode $d$. The $\pi^0 \rightarrow \gamma\gamma$ branching fraction is accounted for in the MC simulation. The resulting values for $\epsilon_{2b}$ and $\epsilon_{3b}$ are $(1.19\pm0.03) \times 10^{-2}$ and $(1.16\pm0.05) \times 10^{-2}$, respectively. The mean is calculated as a generalized weighted mean~\cite{rice2007mathematical}\cite{Cox_2006}, taking into account correlated and uncorrelated uncertainties in a covariance matrix. This approach is used because the difference in systematic uncertainties between the two $\bar{D}^{0}$ decay modes leads to the need to weight them in order to calculate the final branching fraction and uncertainty correctly. Finally, the correction due to DCS decays discussed in Section  \ref{sec:detector} is made.

The results for ${\cal B}$ and ${\cal A}_{CP}$ for $B^{+}\to \bar{D}^{0}\pi^{+}$ are:
\begin{align} \label{eq:BR_Acp_control} 
  {\cal B} & = (4.53 \pm 0.02 \pm 0.15 ) \times 10^{-3}, \\
  {\cal A}_{CP} & = (0.19 \pm 0.36 \pm 0.57)\%. 
\end{align}
The uncertainties quoted are statistical and systematic, respectively. The systematic uncertainties associated with the measurement of ${\cal B}$ and ${\cal A}_{CP}$ are explained in detail in Section \ref{systematics}, and the contributions of each of these are listed in Table \ref{sysYields} and \ref{sysAcp}, respectively. These results are in agreement with the PDG values~\cite{PDG_chi_d} of  ${\cal B} = (4.68 \pm 0.13 ) \times 10^{-3}$ and $ {\cal A}_{CP} = (-0.7 \pm 0.7)\%$.
As a cross-check, we determined ${\cal B}$ and ${\cal A}_{CP}$ for each of the $B_{2b}$ and $B_{3d}$ modes separately, and for just the SVD1 data set. All are in agreement within statistical uncertainties. The fitted yield for each respective category is listed in Table \ref{eff_pi+}.

\begin{table}[htb]
\caption{ Fitted number of signal and backgrounds events for the two reconstruction modes ($B_{2b}$ and $B_{3b}$) of $B^{+} \to \bar{D}^{0} \pi^{+}$. Uncertainties are statistical only.}
\label{eff_pi+}
\begin{tabular}
{@{\hspace{0.3cm}}l@{\hspace{0.3cm}}||@{\hspace{0.3cm}}c@{\hspace{0.3cm}}||@{\hspace{0.3cm}}c@{\hspace{0.3cm}}}
\hline \hline
&&\\[-1em]
Category  & $B_{2b}$ mode ($\times 10^4$) & $B_{3b}$ mode ($\times 10^4$) \\
\hline
 Signal & $4.27\pm0.02$ & $4.18\pm0.02$ \\
 Continuum & $0.70\pm0.01$ & $1.78\pm0.3$ \\
 Generic B & $3.58\pm0.03$ & $3.87\pm0.03$ \\
 Rare & $0.03$ (fixed) & $0.05$ (fixed) \\
\hline \hline
\end{tabular}
\end{table}

\section{\label{sec:results}$B^{0}\to \bar{D}^{0} \pi^{0}$}

\begin{figure*}[htb]
  \includegraphics[width=1.0\textwidth]{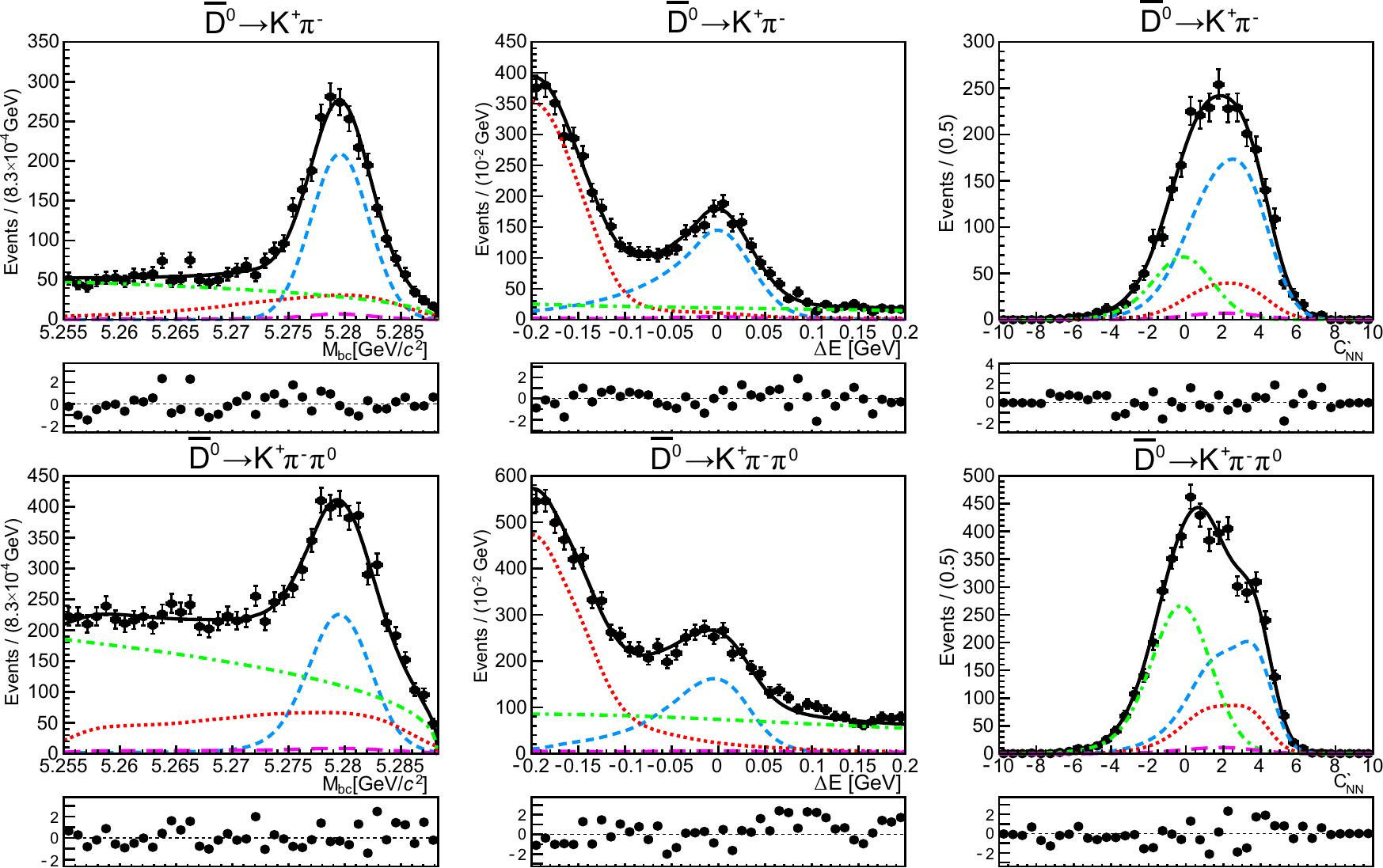}
\caption{Projections of the $B^{0}\to \bar{D}^{0} \pi^{0}$ fit results into the signal region ($5.275<M_{\rm bc} <5.285 \gev$,$-0.12< \Delta E<0.07 \gev$,$-1< C_{\rm NN}' <6$ ) for $M_{\rm bc}$ (left), $\Delta E$ (middle) and $C_{\rm NN}'$ (right) split into the $B_{2b}$ mode (top) and $B_{3b}$ (bottom). The blue short-dashed curve shows the signal PDF, red dotted curve shows the $B\bar{B}$ background PDF, green dash-dotted curve shows the continuum background PDF, pink long-dashed curve shows the (almost negligible) rare background PDF, black line is the fit result. Also shown underneath each graph is the residual pulls between the data points and fitted PDF.}
\label{yield}
\end{figure*}

\begin{figure*}[htb]
\includegraphics[width=0.9\textwidth]{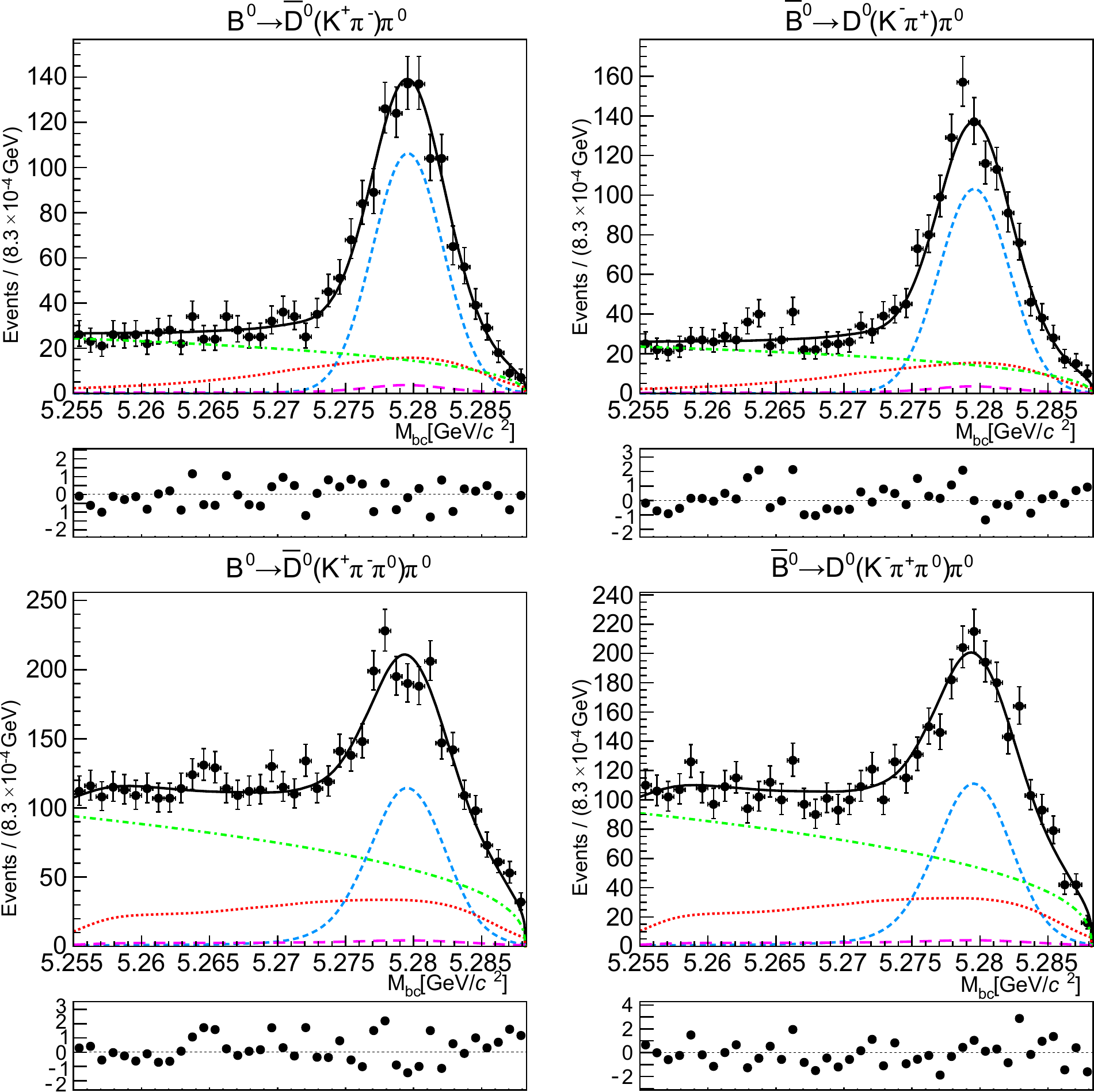}
\caption{Projections of the $B^{0}\to \bar{D}^{0} \pi^{0}$ fit results for $M_{\rm bc}$ into the signal region ($-0.12< \Delta E<0.07 \gev$,$-1< C_{\rm NN}' <6$ ) for $B^{0}\to\bar{D}^{0}(K^{+}\pi^{-})\pi^{0}$ (top left), $\bar{B}^{0}\to D^{0}(K^{-}\pi^{+})\pi^{0}$ (top right), $B^{0}\to\bar{D}^{0}(K^{+}\pi^{-}\pi^{0})\pi^{0}$ (bottom left), $\bar{B}^{0}\to D^{0}(K^{-}\pi^{+} \pi^{0})\pi^{0}$ (bottom right). The blue short-dashed curve shows the signal PDF, red dotted curve shows the $B\bar{B}$ background PDF, green dash-dotted curve shows the continuum background PDF, pink long-dashed curve shows the (almost negligible) rare background PDF, black line is the fit result. Also shown underneath each graph is the residual pulls between the data points and fitted PDF. }
\label{resultsAcp}
\end{figure*}

After applying the calibration factors determined from studies of the $B^{+}\to\bar{D}^{0} \pi^{+}$ mode to the PDFs, we fit the signal $B^{0}\to \bar{D}^{0} \pi^{0}$ PDFs to data and find $4448 \pm 97$ signal events and ${\cal A}_{CP,\text{raw}} = (1.48 \pm 2.05 )\%$. The uncertainties quoted are statistical. As was the case for $B^{+}\to\bar{D}^{0} \pi^{+}$, ${\cal A}_{CP,\text{raw}}$ is the value returned from the fit without a correction for sources of bias. Figure~\ref{yield} shows the signal-enhanced projections of the fits. Figure \ref{resultsAcp} shows signal-enhanced projections of $M_{\rm bc}$ separated into $B^{0}$ and $\bar{B}^{0}$ decays.

Using Eq.~\ref{eq:Br_recipe}, the PDG value ${\cal B} (\Upsilon(4S) \to \bar{B}^{0}B^{0})=(48.6\pm0.6\%)$\cite{PDG_chi_d}, the fraction of signal events in data set $d$, $f^s_{2b} = 0.45$, $f^s_{3b} = 0.55$, and the efficiencies $\epsilon_{2b}=(1.00\pm0.03) \times 10^{-2}$ and $\epsilon_{3b}=(1.21\pm0.07) \times 10^{-2}$, we determine the branching fraction to be:
\begin{equation}
\label{eq:BR}
 {\cal B} (B^{0} \to \bar{D}^{0}\pi^{0}) = (2.70 \pm 0.06 \pm 0.10) \times 10^{-4},
\end{equation}
where the quoted uncertainties are statistical and systematic, respectively. The fitted yield for each respective category is listed in Table \ref{eff}.

\begin{table}[htb]
\caption{ Fitted number of signal and backgrounds events for the two reconstruction modes ($B_{2b}$ and $B_{3b}$) of $B^{0}\to \bar{D}^{0} \pi^{0}$. Uncertainties are statistical only.}
\label{eff}
\begin{tabular}
{@{\hspace{0.3cm}}l@{\hspace{0.3cm}}||@{\hspace{0.3cm}}c@{\hspace{0.3cm}}||@{\hspace{0.3cm}}c@{\hspace{0.3cm}}}
\hline \hline
&&\\[-1em]
Category  & $B_{2b}$ mode ($\times 10^3$) & $B_{3b}$ mode ($\times 10^3$) \\
\hline
 Signal & $2.01\pm0.04$ & $2.44\pm0.05$ \\
 Continuum & $4.26\pm0.06$ & $16.47\pm0.22$ \\
 Generic B & $4.76\pm0.10$ & $8.39\pm0.18$ \\
 Rare & $0.15$ (fixed) & $0.47$ (fixed) \\
\hline \hline
\end{tabular}
\end{table}

The ${\cal A}_{CP}$ correction for the $B^{0}\to \bar{D}^{0} \pi^{0}$ decay is measured in the same way as for the $B^{+}\to\bar{D}^{0} \pi^{+}$ mode. A sideband region of data is defined as $0.1 \gev < \Delta E < 0.4 \gev $ and $5.255\gevcc < M_{\rm bc} < 5.27 \gevcc$. Events in this region consist almost entirely of continuum with an expected ${\cal A}_{CP}$ of zero.
In this region we find ${\cal A}_{CP,\text{sideband}}= (1.02 \pm 0.64)\%$, and we subtract this value and the fit bias ($0.04\%$) from ${\cal A}_{CP,\text{raw}}$ to correct for detector bias.

The direct $\CP$-violation parameter is thus measured to be:
\begin{equation} \label{eq:Acp}
{\cal A}_{\CP} (B^{0} \to \bar{D}^{0}\pi^{0}) = (0.42 \pm 2.05  \pm 1.22 )\%
\end{equation}
The uncertainties quoted are statistical and systematic, respectively.

\section{ Systematic Uncertainties} \label{systematics}

The systematic uncertainties associated with the measurement of ${\cal B}$ and ${\cal A}_{CP}$ are as follows.
 \begin{itemize}
  \item Number of $B\bar{B}$ pairs: the uncertainty associated with the measured number of $B\bar{B}$ pairs in the full data set collected at Belle is 1.37\%~\cite{B_factory}.
  
  \item ${\cal B}(\Upsilon (4S)\to B^0 \bar{B}^{0})$: uncertainty from the branching fraction ${\cal B} (\Upsilon(4S) \to \bar{B}^{0}B^{0})=(48.6\pm0.6\%)$~\cite{PDG_chi_d}.
  
  \item DCS mode correction: the uncertainty due to the correction for doubly Cabibbo-suppressed decays is 0.01\% for both $B^{0}\to\bar{D}^{0}\pi^{0}$ and $B^{+}\to\bar{D}^{0}\pi^{+}$ (see Section \ref{sec:detector}).
  
  \item Charged track efficiency: the uncertainty associated with a possible difference in efficiency between MC and data for charged-track reconstruction is found to be 0.35\% per track using partially reconstructed $D^{*+}\to D^{0}(\to\pi^{+}\pi^{-}\pi^{0})\pi^{+}$ events~\cite{B_factory}. 
  
  \item $\pi^{0}$ detection efficiency:  the ratio of data to MC efficiency for $\pi^{0}$ reconstruction is based on a study of $\tau^{-}\to\pi^{-}\pi^{0} \nu_{\tau}$~ decays~\cite{Ryu}. This ratio is $(96 \pm 2)\%$ per $\pi^{0}$.
  
  \item MC statistics
in efficiency calculation: uncertainty associated with the reconstruction efficiency is based on the binomial statistics of the MC data set used. This is 0.094\% for $B^{0}\to\bar{D}^{0}(K^{+}\pi^{-})\pi^{0}$, 
  0.18\% for $B^{0}\to\bar{D}^{0}(K^{+}\pi^{-}\pi^{0})\pi^{0}$, 
  0.075\% for $B^{+}\to\bar{D}^{0}(K^{+}\pi^{-})\pi^{+}$, and 
  0.19\% for $B^{+}\to\bar{D}^{0}(K^{+}\pi^{-}\pi^{0})\pi^{+}$.
  
  \item $\bar{D}^{0}$ subdecay branching fraction and $A_{CP}$: from the PDG average~\cite{PDG_chi_d}.
  
  \item PID efficiency: systematic error associated with a small difference in PID efficiency between MC and data. This is based on an inclusive  $D^{*+} \to D^{0}(K^{-}\pi^{+})\pi^{+}$ study~\cite{B_factory}. 
  The uncertainty is calculated as 1.3\% for $B^{0}\to\bar{D}^{0}(K^{+}\pi^{-})\pi^{0}$, 
  1.3\% for $B^{0}\to\bar{D}^{0}(K^{+}\pi^{-}\pi^{0})\pi^{0}$, 
  2.2\% for $B^{+}\to\bar{D}^{0}(K^{+}\pi^{-})\pi^{+}$, and 
  2.2\% for $B^{+}\to\bar{D}^{0}(K^{+}\pi^{-}\pi^{0})\pi^{+}$. 
  
  \item Signal decay mode yield ratio $f^{s}_{d}$: the ratio between the $D^{0}$ decay modes in signal, $f^{s}_{d}$, is fixed based on the expected yields from MC. To account for the uncertainty, we perform two fits, varying the fixed value by $\pm1 \sigma$ (based on MC statistics of the simulation). This variation gives changes of $[-0.38, +0.31]$\%	and $[-0.08, +0.19]$\% in $\mathcal{B}$ for $B^{0}\to\bar{D}^{0}\pi^{0}$ and $B^{+}\to\bar{D}^{0}\pi^{+}$ respectively. The uncertainty in ${\cal A}_{CP}$ is $[-0.02, +0.03]$ for $B^{0}\to\bar{D}^{0}\pi^{0}$ and  $<0.01$ for $B^{+}\to\bar{D}^{0}\pi^{+}$. 
  
  \item $C_{\text{NN}}'$ calibration factors: we fit with and without the calibration factors applied to the PDFs. The difference between the yields and ${\cal A}_{CP}$ of these fits is quoted as the uncertainty. The uncertainty in $\mathcal{B}$ is 0.34\% and 0.06\% for $B^{0}\to\bar{D}^{0}\pi^{0}$ and $B^{+}\to\bar{D}^{0}\pi^{+}$, respectively. For ${\cal A}_{CP}$ it is 0.06 and $<0.01$, respectively.
  
  \item Modification of the $B_{3d}$ $M_{\rm bc} \times \Delta E$ KEST PDF: the uncertainty from the $\Delta E$ modification to the $\bar{D}^{0}\to K^{+}\pi^{-}\pi^{0}$ $M_{\rm bc}\times \Delta E$ KEST PDF is evaluated by comparing the fit results obtained using the corrected and uncorrected PDF. The difference in the fitted yields of 0.63\% for $B^{0}\to\bar{D}^{0}\pi^{0}$ and 0.24\% for $B^{+}\to\bar{D}^{0}\pi^{+}$ is quoted as the uncertainty. For ${\cal A}_{CP}$ this is 0.06 and $<0.01$, respectively.
  
  \item 2D KEST PDFs: $B\bar{B}$, rare, and $B_{3b}$ signal $M_{\rm bc} \times \Delta E$ PDFs all use a fixed 2D KEST PDF. To estimate the uncertainty from this, an ensemble test is performed running 1000 fits over the data, with each fit using a different Gaussian-fluctuated KEST PDF based on bin statistics. The uncertainty is quoted as the RMS of the resulting yield and ${\cal A}_{CP}$ distributions. This contributes 0.35\% and 0.05\% to the uncertainty in $\mathcal{B}$ for $B^{0}\to\bar{D}^{0}\pi^{0}$ and $B^{+}\to\bar{D}^{0}\pi^{+}$, respectively. For ${\cal A}_{CP}$ the contribution is 0.15 and $<0.01$, respectively.
  
  \item Fixed rare $B$-decay background yield: the uncertainty due to the rare B yield is the quadratic sum of the statistical uncertainty based on the size of the MC dataset and the uncertainty in the branching fractions used to generate the MC. For modes with three body final states ($K^+\pi^-\pi^0$ and $K^+\pi^-\pi^+$), this latter component is taken from the uncertainty in the PDG branching fractions $\mathcal{B}({B}^{0} \to K^{+} \pi^{-} \pi^{0})=(37.8\pm3.2)\times 10^{-6}$ and $\mathcal{B}({B}^{+} \to K^{+} \pi^{-} \pi^{+})=(51.0\pm2.9)\times 10^{-6}$~\cite{PDG_chi_d}. For $\bar{D}^{0}\to K^{+}\pi^{-}\pi^{0}$ modes, this latter component is taken as the difference between the PDG values for the decays with experimentally measured branching fractions and the branching fractions used in the MC generator (or the uncertainty on the PDG value if that is larger). To estimate the effect on signal yield, the data is refitted, varying the rare yield by $\pm1\sigma$. The uncertainty in $\mathcal{B}$ is 0.47\% for $B^{0}\to\bar{D}^{0}\pi^{0}$ and $0.03$\% for $B^{+}\to\bar{D}^{0}\pi^{+}$.
 
  \item Fit bias: the uncertainty in the fit bias obtained from the signal MC ensemble tests is quoted as an uncertainty. For ${\cal B}$ this is 0.30\% and 0.16\% for $B^{0}\to\bar{D}^{0}\pi^{0}$ and $B^{+}\to\bar{D}^{0}\pi^{+}$, respectively, and for ${\cal A}_{CP}$ it is 0.05 and 0.02, respectively.
  
  \item ${\cal A}_{CP}$ detector bias correction: uncertainty on the correction made to ${\cal A}_{CP}$ is the statistical uncertainty on the  ${\cal A}_{CP,\rm{sideband}}$ measurement, summed in quadrature with the deviation of the ${\cal A}_{CP}$ of the $B^{+}\to\bar{D}^{0} \pi^{+}$ mode from the expected value of ${\cal A}_{CP}=0$. This is 0.66 for $B^{0}\to\bar{D}^{0}\pi^{0}$ and 0.42 for $B^{+}\to\bar{D}^{0}\pi^{+}$.
  
  \item Fixed background ${\cal A}_{CP}$: uncertainties from background ${\cal A}_{CP}$ being fixed in fits are estimated by varying them by $\pm 1 \sigma$ (based on sideband data) and comparing the ${\cal A}_{CP}$ in the resultant fits. This is found to be 0.49 for $B^{0}\to\bar{D}^{0}\pi^{0}$ and 0.03 for $B^{+}\to\bar{D}^{0}\pi^{+}$. This is correlated with the ${\cal A}_{CP}$ detector bias correction uncertainty.
 \end{itemize}

In order to accurately calculate the uncertainty in $\mathcal{B}$, the $\bar{D}^{0}$ decay-mode-dependent factors are combined in a generalized weighted mean as shown in Eq.~\ref{eq:Br_recipe}. The absolute uncertainties for charged track efficiency, $\pi^0$ detection efficiency, reconstruction efficiency, PID efficiency, and $\bar{D}^{0}$ branching fraction are combined into a covariance matrix, $\Sigma$, that accounts for their correlations between the two-body and three-body modes. For $B^{0}\to\bar{D}^{0}\pi^{0}$, $\Sigma = \big[\begin{smallmatrix}
  1.47 & 2.40\\
  2.40 & 6.76
\end{smallmatrix}\big]$, and for $B^{+}\to\bar{D}^{0}\pi^{+}$ $\Sigma = \big[\begin{smallmatrix}
  1.17 & 1.05\\
  1.05 & 4.03
\end{smallmatrix}\big]$. The combined value, which we call ``mean efficiency'', is calculated as $\bar{\epsilon}=\sigma^2_{\bar{\epsilon}} (\mathcal{J}^T\Sigma^{-1}\left[\frac{f^s_{2b}}{\epsilon_{2b}},\frac{f^s_{3b}}{\epsilon_{3b}}\right]^T)^{-1}$, with variance  ${\sigma^2_{\bar{\epsilon}}=(\mathcal{J}^T\Sigma^{-1}\mathcal{J})^{-1}}$ (where $\mathcal{J}=[1,1]^T$)~\cite{rice2007mathematical}\cite{Cox_2006}. The relative uncertainty on this is found to be 2.43\% for $B^{0}\to\bar{D}^{0}\pi^{0}$, and 2.54\% for $B^{+}\to\bar{D}^{0}\pi^{+}$.

\begin{table}[htb]
\caption{Systematic uncertainties for ${\cal B}$ measurements. The mean efficiency results from combining charged track efficiency, $\pi^0$ detection efficiency, MC statistics in efficiency calculation, $\bar{D}^{0}$ subdecay properties, and PID efficiency in a general weighted mean calculation.}
\label{sysYields}
\begin{tabular}
{@{\hspace{0.2cm}}l@{\hspace{0.2cm}}||@{\hspace{0.2cm}}c@{\hspace{0.2cm}}||@{\hspace{0.2cm}}c@{\hspace{0.2cm}}}
\hline \hline
Systematic  & $B^{0}\to\bar{D}^{0}\pi^{0}$ & $B^{+}\to\bar{D}^{0}\pi^{+}$   \\
\hline
No. $B\bar{B}$ &	1.37\% &	1.37\% \\ \hline
${\cal B}(\Upsilon (4S)\to B^{0} \bar{B}^{0})$ &	1.23\% &	1.17\% \\ \hline
DCS mode correction &	0.01\% &	0.01\% \\ \hline
Mean efficiency &	2.43\%	& 2.54\% \\ \hline
&&\\[-1em]
Fixed $f^{s}_{d}$ &	${+0.31 \atop -0.38}$\%	& ${+0.19 \atop -0.08}$\% \\
&&\\[-1em]\hline
Cal. Factors ($C_{\text{NN}}'$) & 0.34\% &0.06\% \\ \hline
$\Delta E$ KEST modification & 0.63\% &	0.24\% \\ \hline
KEST PDFs &	0.35\% & 0.05\%	\\ \hline
Fixed Rare Yields &	0.47\% &	0.03\% \\ \hline
Fit bias &	0.30\%	& 0.16\% \\ \hline
Bkg. ${\cal A}_{CP}$  & 0.01\% & 0.05\% \\ \hline
Total & 3.65\% & 3.32\% \\
\hline \hline
\end{tabular}
\end{table}

\begin{table}[htb]
\caption{Systematic uncertainties for ${\cal A}_{CP}$ measurements. All numbers listed $\times 10^{-2}$. * denotes correlated uncertainties.}
\label{sysAcp}
\begin{tabular}
{@{\hspace{0.2cm}}l@{\hspace{0.2cm}}||@{\hspace{0.2cm}}c@{\hspace{0.2cm}}||@{\hspace{0.2cm}}c@{\hspace{0.2cm}}}
\hline \hline
Systematic for ${\cal A}_{CP}$  & $B^{0}\to\bar{D}^{0}\pi^{0}$ & $B^{+}\to\bar{D}^{0}\pi^{+}$   \\
\hline
$\bar{D}^{0}$ Decay ${\cal A}_{CP}$ & $0.35$ &	$0.35$ \\ \hline
&&\\[-1em]
Fixed $f^{s}_{d}$ &	${+0.03 \atop -0.02}$ & $<0.01$ \\
&&\\[-1em]\hline
Cal. Factors ($C_{\text{NN}}'$) & $0.06$ & $<0.01$ \\ \hline
$\Delta E$ KEST modification & $0.06$ &	$<0.01$ \\ \hline
KEST PDFs &	$0.15$ & $<0.01$ 	\\ \hline
Fixed Rare Yields &	$<0.01$ & $<0.01$ \\ \hline
Fit bias &	$0.05$	& $0.02$ \\ \hline
Detector bias (signal)* &	$0.66$ & $0.42$ \\ \hline
Detector bias (background)* & $0.49$ &	$0.03$ \\ \hline
Total & 1.22 & 0.57 \\
\hline \hline
\end{tabular}
\end{table}

The values of all contributions to the branching fractions are listed in Table \ref{sysYields}. The quadratic sum of these terms is quoted as the total systematic uncertainty for ${\cal B}$.
The values of all contributions to the ${\cal A}_{CP}$ measurements are listed in Table \ref{sysAcp}. The quadratic sum of these terms is quoted as the total systematic uncertainty for ${\cal A}_{CP}$.

\section{\label{sec:conclusions}Conclusions}

Our measurements of 
\begin{align} \label{eq:Br_B0_conc}
\mathcal{B}(B^0\to \bar{D}^{0}\pi^{0}) &= [2.70 \pm 0.06~ \pm 0.10~ ] \times 10^{-4},  \\
{\cal B}(B^+\to \bar{D}^{0}\pi^{+}) &= (4.53 \pm 0.02 \pm 0.15 ) \times 10^{-3}
\end{align}
are the most precise to date. They agree with our previous measurements~\cite{blyth}\cite{Belle_Dpi_Br} within uncertainties, and supersede those results. They are also in agreement with PDG values~\cite{PDG_chi_d}. 

Our result
\begin{equation} \label{eq:Acp_B0_conc}
{\cal A}_{\CP} (B^{0} \to \bar{D}^{0}\pi^{0}) = (0.42 \pm 2.05  \pm 1.22 )\%  \\
\end{equation}
is the first reported for this mode. Our result
\begin{equation} \label{eq:Acp_B+_conc}
{\cal A}_{CP}(B^+\to \bar{D}^{0}\pi^{+}) = (0.19 \pm 0.36 \pm 0.57)\%  \\
\end{equation}
is the most precisely measured and agrees with our previous result~\cite{Belle_Dpi_Acp}, which it supersedes.

%
%

\section{\label{acknowledgements}Acknowledgments}

We thank the KEKB group for the excellent operation of the
accelerator; the KEK cryogenics group for the efficient
operation of the solenoid; and the KEK computer group, and the Pacific Northwest National
Laboratory (PNNL) Environmental Molecular Sciences Laboratory (EMSL)
computing group for strong computing support; and the National
Institute of Informatics, and Science Information NETwork 5 (SINET5) for
valuable network support.  We acknowledge support from
the Ministry of Education, Culture, Sports, Science, and
Technology (MEXT) of Japan, the Japan Society for the 
Promotion of Science (JSPS), and the Tau-Lepton Physics 
Research Center of Nagoya University; 
the Australian Research Council including grants
DP180102629, 
DP170102389, 
DP170102204, 
DP150103061, 
FT130100303; 
Austrian Federal Ministry of Education, Science and Research (FWF) and
FWF Austrian Science Fund No.~P~31361-N36;
the National Natural Science Foundation of China under Contracts
No.~11435013,  
No.~11475187,  
No.~11521505,  
No.~11575017,  
No.~11675166,  
No.~11705209;  
Key Research Program of Frontier Sciences, Chinese Academy of Sciences (CAS), Grant No.~QYZDJ-SSW-SLH011; 
the  CAS Center for Excellence in Particle Physics (CCEPP); 
the Shanghai Science and Technology Committee (STCSM) under Grant No.~19ZR1403000; 
the Ministry of Education, Youth and Sports of the Czech
Republic under Contract No.~LTT17020;
Horizon 2020 ERC Advanced Grant No.~884719 and ERC Starting Grant No.~947006 ``InterLeptons'' (European Union);
the Carl Zeiss Foundation, the Deutsche Forschungsgemeinschaft, the
Excellence Cluster Universe, and the VolkswagenStiftung;
the Department of Atomic Energy (Project Identification No. RTI 4002) and the Department of Science and Technology of India; 
the Istituto Nazionale di Fisica Nucleare of Italy; 
National Research Foundation (NRF) of Korea Grant
Nos.~2016R1\-D1A1B\-01010135, 2016R1\-D1A1B\-02012900, 2018R1\-A2B\-3003643,
2018R1\-A6A1A\-06024970, 2019K1\-A3A7A\-09033840,
2019R1\-I1A3A\-01058933, 2021R1\-A6A1A\-03043957,
2021R1\-F1A\-1060423, 2021R1\-F1A\-1064008;
Radiation Science Research Institute, Foreign Large-size Research Facility Application Supporting project, the Global Science Experimental Data Hub Center of the Korea Institute of Science and Technology Information and KREONET/GLORIAD;
the Polish Ministry of Science and Higher Education and 
the National Science Center;
the Ministry of Science and Higher Education of the Russian Federation, Agreement 14.W03.31.0026, 
and the HSE University Basic Research Program, Moscow; 
University of Tabuk research grants
S-1440-0321, S-0256-1438, and S-0280-1439 (Saudi Arabia);
the Slovenian Research Agency Grant Nos. J1-9124 and P1-0135;
Ikerbasque, Basque Foundation for Science, Spain;
the Swiss National Science Foundation; 
the Ministry of Education and the Ministry of Science and Technology of Taiwan;
and the United States Department of Energy and the National Science Foundation.

\bibliography{PRD_B_D0pi0_v01}

\begin{thebibliography}{38}
\bibitem [{CC()}]{CC}%
  \BibitemOpen
  \href@noop {} {}\bibinfo {note} {Throughout this Letter, the inclusion of the
  charge-conjugate decay modes is implied unless stated otherwise.}\BibitemShut
  {Stop}%
\bibitem [{alp()}]{alpha}%
  \BibitemOpen
  \href@noop {} {}\bibinfo {note} {Another naming convention, $\beta$ ($=
  \phi_1$), $\alpha$ ($= \phi_2$) and $\gamma$ ($=\phi_3$) is also used in the
  literature.}\BibitemShut {Stop}%
\end{thebibliography}

\end{document}